\documentclass[aps,prb,twocolumn]{revtex4}
\usepackage{xcolor}

\def\be{\begin{equation}}       \def\ee{\end{equation}}
\def\bea{\begin{eqnarray}}      \def\eea{\end{eqnarray}}
\def\ba{\begin{array} }
\def\ea{\end{array} }
\def\bnum{\begin{enumerate} }
\def\enum{\end{enumerate}}

\def\=>{\Rightarrow}
\def\>{\rightarrow}

\def\eye2{Fathbb{I}}

\def\d0{\Delta_{0}}

\usepackage{amsmath}

\begin{document}
\title{\bf 
Linking the pseudo-gap in the cuprates with local symmetry breaking: a commentary}
\author{S. A. Kivelson}
\affiliation{Department of Physics, Stanford University, Stanford, CA 94305, USA}
\author{Samuel Lederer}
\affiliation{Department of Physics, Cornell University, USA}

\date{\today }

\maketitle

In the last two decades, increasingly precise measurements have established that the cuprate high temperature superconductors exhibit 
 numerous ordering tendencies.
In addition to the ``big two''  -- N{\'e}el antiferromagnetism and d-wave superconductivity (SC) -- there are a variety of other orders  that have been observed, especially in the enigmatic pseudogap regime of the phase diagram.  The term ``pseudogap'' denotes a suppression of the 
density of states between a  doping dependent crossover temperature, $T^*(p)$, and  the (lower) SC transition temperature, $T_c(p)$.
Thus, for doping less than $p^*\approx 0.19$ (above which $T^*(p)$ vanishes), the pseudogap is the ``normal state" out of which superconductivity emerges. 

Mukhopadhyay {\it et al}\cite{mukhopadhyay2019}, in a paper in the current issue of PNAS, venture a bold proposition as to the microscopic origin of the pseudogap on the basis of a careful examination of high resolution scanning tunneling spectroscopy (STS) on the material Bi$_2$Sr$_2$CaCu$_2$O$_{8+x}$ (Bi-2212). 
They focus on two distinct forms of order:  charge density wave (CDW),  
the breaking of lattice translation symmetry; and nematicity, 
the breaking of lattice rotational symmetry.  
Both of these ordering tendencies have been 
identified across all families of hole-doped cuprates.
By measuring tunneling conductance with atomic resolution over a large field of view, performing a Fourier transform, and analyzing data from distinct regions of momentum space, 
Mukhopadhyay {\it et al} 
identify energies $\Delta^*$, $E^D_{max}$, and $E^N_{max}$ that characterize the pseudogap, CDW, and nematicity respectively. Measured on samples whose doping spans 
regime, these energies are, within experimental error, identical.

On the basis of this remarkable result, the authors argue that the pseudogap is a consequence of the tendency towards a unidirectional density wave that, if long range ordered, 
would break both translation and rotation symmetry. 
In the presence of disorder, translation symmetry breaking cannot occur (unless the CDW is commensurately locked to the lattice).
However,
a phase with ``vestigial"  nematic order, i.e. rotational symmetry breaking, survives to a critical disorder strength.
\cite{Nie2014,kivelson-1998}
 The transition temperature for this nematic order would then provide a sharp definition for $T^*$. 
 In brief, the proposition is:

\bigskip
\noindent{\emph{The pseudogap is due 
to density wave correlations rendered short-range by disorder. These short-range correlations produce a phase with vestigial nematic order, whose transition temperature  
determines $T^*$.}}

\bigskip

While the 
arguments leading to this conclusion are highly suggestive, there are a number of important subtleties and challenges that still remain to be addressed.  In the remainder of this commentary, we discuss some of the most significant -- not as a critique 
but as a road map for further investigation.

\section{How definitive is the spectroscopic evidence?}   To establish the link between CDW and nematic orders and  the pseudogap, the authors compare three separate spectroscopic measurements.  Each relies on data from 
non-overlapping regimes of Fourier space:  the CDW spectrum, $D^Z({\bf Q},E)$ from near the ordering vectors ${\bf Q}={\bf Q}_x^D \approx (0.25,0)$ and ${\bf Q}={\bf Q}_y^D\approx (0,0.25)$, the nematic $N^Z(E)$ from the reciprocal lattice vectors of the crystal ${\bf Q}={\bf Q}_x^B=(1,0)$ and ${\bf Q}={\bf Q}_y^B=(0,1)$ , and the averaged (${\bf Q}=\bf 0$) density of states $\rho(E)$. 
However, all are based on single-particle spectra, and so to some extent are sensitive to the density of states.  We thus have to ask whether the observed coincidence of the maxima in the CDW and nematic spectra with $\Delta^\star$ might be less meaningful than it 
seems.  We do not  know a quantitative way to address this issue.  However, while the drop of all three of these quantities with decreasing $E$ for $E < \Delta^\star$ might be simply a density of states effect, 
both $N^Z(E)$ and $D^Z({\bf Q},E)$ drop much more rapidly with increasing $E> \Delta^\star$ than does $\rho(E)$.  This makes it difficult to avoid the conclusion that these orders are 
tied to the pseudogap.

\section{Other orders:} Just as nematic order can serve as an avatar for CDW order, there may 
be other ordering tendencies for which the presence of one can serve as indirect evidence of another. 
 Indeed, the authors note that the CDW could be a subsidiary order,
 reflecting a primary tendency toward a pair density wave (PDW). 
 While recent experiments\cite{blackburn2019,abbamonte}
 make the ``PDW as mother of all orders'' proposition less likely, 
  the notion that the pseudogap 
 involves local SC pairing is both theoretically plausible and supported by a variety of 
  experiments
  \cite{carlson2004,wang2001,gomes2007,loew2019,Mankowsky2017,zhou2019}. Moreover,
  at very low doping, 
  the pseudogap is associated with the growth of the AF correlation length beyond a few 
  lattice constants \cite{keimer1992,Gull2015,scheurer2018,huang2019,loew2019},
 although the AF correlation length is 
   less than a lattice constant in the pseudogap regime of near-optimally doped  YBa$_2$Cu$_3$O$_{7-x}$ (YBCO) and Bi-2212\cite{Xu2009}.
Accordingly, the pseudogap may not have just one cause. That said, we applaud the authors for the clarity of their conjecture; ``all of the above" is an 
unsatisfying (though possibly technically correct) answer to the question of the pseudogap. Indeed, exploring the validity of the proposed unifying perspective should stimulate future experiments. For instance, the suggested relation between nematicity and $T^*$ implies that uniaxial strain should be a fruitful knob to turn.

In the context of the authors' proposal that the pseudogap is due to density wave fluctuations, it important to bear in mind that ascribing a spectroscopic pseudogap to fluctuations of an order parameter is an inherently imprecise notion.  
 A system illustrating this 
 point\cite{carlson2004} is a one dimensional 
 system with a Luther-Emery liquid ground state.  At 
 high temperatures, the 
system behaves as a Luttinger liquid, with gapless charge and spin excitations. 
At a crossover temperature $T^\star$, 
a spin gap opens.  For 
$T<T^\star$, both CDW and SC correlations grow strongly. 
An array of such systems weakly coupled together will order at some temperature 
$T_o\ll T^*$, but whether as a SC or CDW state depends on 
a variety of details. 
This is a solved problem 
without a straightforward intuitive understanding
Viewing the spin gap (essentially a pseudogap) as 
due to CDW fluctuations is reasonable, but an equally good case 
exists for SC fluctuations. 
And neither 
perspective fully captures the underlying physics.

\section{Ambiguity concerning the $\vec Q = \vec 0$ order:} 
The $T^\star$ line in the central phase diagram presented by Mukhopadhyay {\it et al}  shows points at which various probes have provided evidence 
for the uniform (${\bf Q}={\bf 0}$) breaking of a symmetry. 
(Several other studies merit inclusion here, such as Refs. \cite{hinkov2008,Daou2009,wu2015}).
These experiments provide evidence that numerous symmetries are broken at $T^*$ (a logical possibility, though one requiring fine tuning). If so, 
the authors' 
proposal that $T^*$ is a nematic transition is incomplete. 
Of course, it may eventually turn out 
that all these  experiments are detecting 
the same transition, one of primarily nematic character.

\section{Is  $T^*$ unique?} For the most part, the experimental studies of local CDW order have identified an onset temperature, $T_{cdw}$, that is lower than $T^*$,  and generally decreases with decreasing  $p <1/8$ while $T^*$ increases.  (See, for example, Ref. \cite{parker2010,cyrchoiniere2015}).  Mukhopadhyay {\it et al} address this, quite reasonably,  by noting that the onset of CDW order is never sharp, so that it may merely mark the temperature at which a signal becomes detectable above background, and  hence may  not be  physically meaningful.

On the other hand, more than one analysis of the crossover phenomena observed in the pseudogap regime has identified a distinct crossover temperature
$T^{**} \approx T_{cdw} < T^*$  (e.g., see Refs. \cite{emery-1997} and \cite{tranquada-2019_2}). In addition to the CDW onset, several relatively well defined and physically significant experimental signals follow $T^{**}$ rather than $T^*$, such as the onset of an unusual Nernst effect and weak diamagnetism, as well as a maximum in the NMR relaxation rate $1/T_1T$  -- phenomena often associated with the growth of SC correlations. Such observations  suggest that $T^{**}$ may be a physically meaningful temperature scale and not just a detection threshold. 

Even if true, this may be an unimportant subtlety of the crossover physics:  Given that there is no long range order of either magnetism or SC, 
and that no sharp transition to a density wave 
ordered state is likely to survive in the presence of disorder, identifying $T^*$ with a 
nematic transition might offer the only precise way to characterize the pseudogap regime.  Even if this is the case, identifying the extent of significant SC fluctuations above $T_c$ is an important issue in its own right, whether or not it is entirely correlated with $T^*$.

\section{Material specific differences }

One of the appeals of the present proposal is its 
universality.  Because the building blocks of the high $T_c$ cuprates are
 similar nearly square Cu-O planes, it is
 generally accepted that the {\it essential} physics is the same for all ``families'' of these materials.  However, many properties, including some directly relevant to the authors' proposal, differ substantially between families
 , and an overarching understand may require incorporating this diversity. Specifically:
  
 \vspace{\baselineskip} 
\noindent \textbf{A:} The CDW 
signal central to the present discussion (measured by STS on Bi-2212 in zero magnetic field)
has a pronounced d-wave form factor.  However, 
X-ray scattering studies of the 214 family of cuprates such as La$_{2-x}$Sr$_x$CuO$_2$(LSCO) indicate a predominantly s-wave form factor in these materials\cite{achkar2013}, while in in YBCO the evidence is currently inconclusive\cite{mcmahon2019}.
Moreover, the field-induced CDW order seen in STS in the vortex cores of Bi-2212\cite{edkins2019} also has a dominantly s-wave form factor. 
These form factors do not 
correspond to order parameter symmetries, as the CDW ordering vector itself breaks the point group symmetry.  Thus, it is possible that these  differences in form factor are not essential pieces of  the physics.  Alternatively, it may be that 
multiple distinct types of CDW order are present, possibly with the s-wave form factor component in vortex cores being generated as the second harmonic of a PDW\cite{agterberg2019}.

\vspace{\baselineskip} \noindent \textbf{B:} 
The CDW wavelength is not universal. The sign of its doping dependence varies among families, and its value ranges from roughly three lattice constants (as in YBCO) to four (as in Bi-2212). The CDW may even be commensurate in some materials\cite{Zhang2019}, in principle permitting long range order.
\vspace{\baselineskip}\noindent  
\textbf{C:} Spin density wave (SDW) order is often observed, and its interplay with CDW 
strongly depends on the material. 
For instance, in LSCO the CDW and SDW orders
are mutually commensurate and appear to cooperate: wherever CDW 
is observed, SDW 
appears at a 
lower temperature.  In contrast, in YBCO, the 
SDW and CDW 
are mutually incommensurate, and 
apparently compete so ferociously that they never coexist.
 
\vspace{\baselineskip}\noindent\textbf{D:} The proposed scenario involves a significant role for quenched disorder in producing a vestigial nematic phase.  In practice, all the cuprates (with the possible exception of Y$_2$Ba$_4$Cu$_8$O$_{16}$) form non-stoichiometric crystals, so 
some disorder is unavoidable. However, the character of the disorder varies between families of materials.  This might give rise  to  material specific aspects to the coupling between disorder and CDW order. 

\vspace{\baselineskip}
These complexities should not be barriers to a single synthetic perspective, but may be crucial when applying this perspective to individual materials.

\section{ Is the CDW order ``strong'' enough?}  
 Since CDW order has been observed by many different experimental probes, its presence 
 in the pseudogap regime is uncontroversial.  It also has been 
 established from numerical studies that it is one of the leading ordering tendencies of paradigmatic models studied in this context, such as the Hubbard\cite{zheng2017} and $t-J$\cite{corboz2014} models.  Moreover, the CDW order is strong enough to significantly suppress $T_c$ under certain circumstances\cite{cyrchoiniere2018}.

 However, CDW order does not appear as strongly as in more conventional CDW materials, such as the rare-earth tritellurides, RTe$_3$\cite{Brouet2008}. In comparison with RTe$_3$, the CDW ordering peaks observed in hard X-ray diffraction studies of the cuprates are several orders of magnitude weaker, and signatures of band reconstruction in angle-resolved photoemission (ARPES) are, at best, much more subtle\cite{Matt2015}. Thus, there is a question whether the CDW order in the cuprates is ``strong'' enough to account for the pseudo-gap.  
 
 It is not clear, even in principle, how to quantify this issue.  In contrast to the above evidence of weakness, the modulations in the local density of states observed in STM are large
 --order one effects.
 Moreover, estimates from NMR\cite{Wu2011} yield charge density variations of order $0.03$ $e$ per Cu atom, which is substantial.
 
 \vspace{\baselineskip}
\section{Perspective:}  
 High temperature superconductivity was discovered in the cuprates more than thirty years ago. Initially, it was thought that it would admit an elegant ``solution'' - although there was considerable disagreement about whose proposed solution that would 
be.  Since then, through a combination of remarkable advances in material perfection, experimental probes
, and computational methods, we have uncovered a plethora of phenomena of interest in their own right, and which 
reveal the complexity of the problems at hand.  Perhaps the most significant aspect of the Mukhopadhyay {\it et al} paper is that it refocuses attention on the big questions. We have raised above a number of issues to be reconciled with their proposition. However, in such a complex system, the failure of a theory to
 account for some observed behaviors -- so long as they are in some sense ``inessential''--is a shortcoming that is expected and should be tolerated\cite{Kivelson2018}.

\acknowledgments{We thank M-H. Julien, J. Tranquada, S. Hayden, J.C. Seamus Davis, and D. Hawthorn, 
for useful discussions.  SAK supported in part by the U. S. Department of Energy (DOE) Office of Basic Energy Science, at Stanford under contract No. DE-AC02-76SF00515. SL supported in part by a Bethe/KIC fellowship.}


\begin{thebibliography}{35}
\expandafter\ifx\csname natexlab\endcsname\relax\def\natexlab#1{#1}\fi
\expandafter\ifx\csname bibnamefont\endcsname\relax
  \def\bibnamefont#1{#1}\fi
\expandafter\ifx\csname bibfnamefont\endcsname\relax
  \def\bibfnamefont#1{#1}\fi
\expandafter\ifx\csname citenamefont\endcsname\relax
  \def\citenamefont#1{#1}\fi
\expandafter\ifx\csname url\endcsname\relax
  \def\url#1{\texttt{#1}}\fi
\expandafter\ifx\csname urlprefix\endcsname\relax\def\urlprefix{URL }\fi
\providecommand{\bibinfo}[2]{#2}
\providecommand{\eprint}[2][]{\url{#2}}

\bibitem[{\citenamefont{Mukhopadhyay et~al.}(2019)\citenamefont{Mukhopadhyay,
  Sharma, Kim, Edkins, Hamidian, Eisaki, Uchida, Kim, Lawler, Mackenzie
  et~al.}}]{mukhopadhyay2019}
\bibinfo{author}{\bibfnamefont{S.}~\bibnamefont{Mukhopadhyay}},
  \bibinfo{author}{\bibfnamefont{R.}~\bibnamefont{Sharma}},
  \bibinfo{author}{\bibfnamefont{C.~K.} \bibnamefont{Kim}},
  \bibinfo{author}{\bibfnamefont{S.~D.} \bibnamefont{Edkins}},
  \bibinfo{author}{\bibfnamefont{M.~H.} \bibnamefont{Hamidian}},
  \bibinfo{author}{\bibfnamefont{H.}~\bibnamefont{Eisaki}},
  \bibinfo{author}{\bibfnamefont{S.-i.} \bibnamefont{Uchida}},
  \bibinfo{author}{\bibfnamefont{E.-A.} \bibnamefont{Kim}},
  \bibinfo{author}{\bibfnamefont{M.~J.} \bibnamefont{Lawler}},
  \bibinfo{author}{\bibfnamefont{A.~P.} \bibnamefont{Mackenzie}},
  \bibnamefont{et~al.}, \bibinfo{journal}{PNAS} \textbf{\bibinfo{volume}{116}}, \bibinfo{pages}{13249}
  (\bibinfo{year}{2019}).

\bibitem[{\citenamefont{Nie et~al.}(2014)\citenamefont{Nie, Tarjus, and
  Kivelson}}]{Nie2014}
\bibinfo{author}{\bibfnamefont{L.}~\bibnamefont{Nie}},
  \bibinfo{author}{\bibfnamefont{G.}~\bibnamefont{Tarjus}}, \bibnamefont{and}
  \bibinfo{author}{\bibfnamefont{S.~A.} \bibnamefont{Kivelson}},
  \bibinfo{journal}{PNAS}
  \textbf{\bibinfo{volume}{111}}, \bibinfo{pages}{7980} (\bibinfo{year}{2014}).

\bibitem[{\citenamefont{Kivelson et~al.}(1998)\citenamefont{Kivelson, Fradkin,
  and Emery}}]{kivelson-1998}
\bibinfo{author}{\bibfnamefont{S.~A.} \bibnamefont{Kivelson}},
  \bibinfo{author}{\bibfnamefont{E.}~\bibnamefont{Fradkin}}, \bibnamefont{and}
  \bibinfo{author}{\bibfnamefont{V.~J.} \bibnamefont{Emery}},
  \bibinfo{journal}{Nature} \textbf{\bibinfo{volume}{393}},
  \bibinfo{pages}{550} (\bibinfo{year}{1998}).

\bibitem[{\citenamefont{Blackburn~\emph{et al.}}(2019)}]{blackburn2019}
\bibinfo{author}{\bibfnamefont{E.}~\bibnamefont{Blackburn~\emph{et al.}}}
  (\bibinfo{year}{2019}), \bibinfo{note}{manuscript in preparation}.

\bibitem[{\citenamefont{Abbamonte}()}]{abbamonte}
\bibinfo{author}{\bibfnamefont{P.}~\bibnamefont{Abbamonte}},
  \bibinfo{note}{private communication}.

\bibitem[{\citenamefont{Carlson et~al.}(2004)\citenamefont{Carlson, Emery,
  Kivelson, and Orgad}}]{carlson2004}
\bibinfo{author}{\bibfnamefont{E.~W.} \bibnamefont{Carlson}},
  \bibinfo{author}{\bibfnamefont{V.~J.} \bibnamefont{Emery}},
  \bibinfo{author}{\bibfnamefont{S.~A.} \bibnamefont{Kivelson}},
  \bibnamefont{and} \bibinfo{author}{\bibfnamefont{D.}~\bibnamefont{Orgad}}, in
  \emph{\bibinfo{booktitle}{The Physics of Conventional and Unconventional
  Superconductors}}, edited by \bibinfo{editor}{\bibfnamefont{K.~H.}
  \bibnamefont{Bennemann}} \bibnamefont{and}
  \bibinfo{editor}{\bibfnamefont{J.~B.} \bibnamefont{Ketterson}}
  (\bibinfo{publisher}{Springer-Verlag}, \bibinfo{address}{Berlin},
  \bibinfo{year}{2004}), vol.~\bibinfo{volume}{II}.

\bibitem[{\citenamefont{Wang and Ong}(2001)}]{wang2001}
\bibinfo{author}{\bibfnamefont{Y.}~\bibnamefont{Wang}} \bibnamefont{and}
  \bibinfo{author}{\bibfnamefont{N.~P.} \bibnamefont{Ong}},
  \bibinfo{journal}{PNAS}
  \textbf{\bibinfo{volume}{98}}, \bibinfo{pages}{11091} (\bibinfo{year}{2001}).

\bibitem[{\citenamefont{Gomes et~al.}(2007)\citenamefont{Gomes, Pasupathy,
  Pushp, Ono, Ando, and Yazdani}}]{gomes2007}
\bibinfo{author}{\bibfnamefont{K.~K.} \bibnamefont{Gomes}},
  \bibinfo{author}{\bibfnamefont{A.~N.} \bibnamefont{Pasupathy}},
  \bibinfo{author}{\bibfnamefont{A.}~\bibnamefont{Pushp}},
  \bibinfo{author}{\bibfnamefont{S.}~\bibnamefont{Ono}},
  \bibinfo{author}{\bibfnamefont{Y.}~\bibnamefont{Ando}}, \bibnamefont{and}
  \bibinfo{author}{\bibfnamefont{A.}~\bibnamefont{Yazdani}},
  \bibinfo{journal}{Nature} \textbf{\bibinfo{volume}{447}},
  \bibinfo{pages}{569} (\bibinfo{year}{2007}).

\bibitem[{\citenamefont{Loew and Keimer}()}]{loew2019}
\bibinfo{author}{\bibfnamefont{T.}~\bibnamefont{Loew}} \bibnamefont{and}
  \bibinfo{author}{\bibfnamefont{B.}~\bibnamefont{Keimer}},
  \bibinfo{note}{unpublished}.

\bibitem[{\citenamefont{{Mankowsky} et~al.}(2017)\citenamefont{{Mankowsky},
  {Fechner}, {F{\"o}rst}, {von Hoegen}, {Porras}, {Loew}, {Dakovski},
  {Seaberg}, {M{\"o}ller}, {Coslovich} et~al.}}]{Mankowsky2017}
\bibinfo{author}{\bibfnamefont{R.}~\bibnamefont{{Mankowsky}}},
  \bibinfo{author}{\bibfnamefont{M.}~\bibnamefont{{Fechner}}},
  \bibinfo{author}{\bibfnamefont{M.}~\bibnamefont{{F{\"o}rst}}},
  \bibinfo{author}{\bibfnamefont{A.}~\bibnamefont{{von Hoegen}}},
  \bibinfo{author}{\bibfnamefont{J.}~\bibnamefont{{Porras}}},
  \bibinfo{author}{\bibfnamefont{T.}~\bibnamefont{{Loew}}},
  \bibinfo{author}{\bibfnamefont{G.~L.} \bibnamefont{{Dakovski}}},
  \bibinfo{author}{\bibfnamefont{M.}~\bibnamefont{{Seaberg}}},
  \bibinfo{author}{\bibfnamefont{S.}~\bibnamefont{{M{\"o}ller}}},
  \bibinfo{author}{\bibfnamefont{G.}~\bibnamefont{{Coslovich}}},
  \bibnamefont{et~al.},
  \bibinfo{eid}{arXiv:1701.08358} (\bibinfo{year}{2017}).

\bibitem[{\citenamefont{Zhou et~al.}(2019)\citenamefont{Zhou, Chen, Liu, abd
  Anthony T.~Bollinger, He, Bozovic, and Natelson}}]{zhou2019}
\bibinfo{author}{\bibfnamefont{P.}~\bibnamefont{Zhou}},
  \bibinfo{author}{\bibfnamefont{L.}~\bibnamefont{Chen}},
  \bibinfo{author}{\bibfnamefont{Y.}~\bibnamefont{Liu}},
  \bibinfo{author}{\bibfnamefont{I.~S.} \bibnamefont{abd Anthony
  T.~Bollinger}}, \bibinfo{author}{\bibfnamefont{X.}~\bibnamefont{He}},
  \bibinfo{author}{\bibfnamefont{I.}~\bibnamefont{Bozovic}}, \bibnamefont{and}
  \bibinfo{author}{\bibfnamefont{D.}~\bibnamefont{Natelson}}
  (\bibinfo{year}{2019}), \bibinfo{note}{unpublished}.

\bibitem[{\citenamefont{Keimer et~al.}(1992)\citenamefont{Keimer, Belk,
  Birgeneau, Cassanho, Chen, Greven, Kastner, Aharony, Endoh, Erwin
  et~al.}}]{keimer1992}
\bibinfo{author}{\bibfnamefont{B.}~\bibnamefont{Keimer}},
  \bibinfo{author}{\bibfnamefont{N.}~\bibnamefont{Belk}},
  \bibinfo{author}{\bibfnamefont{R.~J.} \bibnamefont{Birgeneau}},
  \bibinfo{author}{\bibfnamefont{A.}~\bibnamefont{Cassanho}},
  \bibinfo{author}{\bibfnamefont{C.~Y.} \bibnamefont{Chen}},
  \bibinfo{author}{\bibfnamefont{M.}~\bibnamefont{Greven}},
  \bibinfo{author}{\bibfnamefont{M.~A.} \bibnamefont{Kastner}},
  \bibinfo{author}{\bibfnamefont{A.}~\bibnamefont{Aharony}},
  \bibinfo{author}{\bibfnamefont{Y.}~\bibnamefont{Endoh}},
  \bibinfo{author}{\bibfnamefont{R.~W.} \bibnamefont{Erwin}},
  \bibnamefont{et~al.}, \bibinfo{journal}{Phys. Rev. B}
  \textbf{\bibinfo{volume}{46}}, \bibinfo{pages}{14034} (\bibinfo{year}{1992}).

\bibitem[{\citenamefont{Gull and Millis}(2015)}]{Gull2015}
\bibinfo{author}{\bibfnamefont{E.}~\bibnamefont{Gull}} \bibnamefont{and}
  \bibinfo{author}{\bibfnamefont{A.~J.} \bibnamefont{Millis}},
  \bibinfo{journal}{Phys. Rev. B} \textbf{\bibinfo{volume}{91}},
  \bibinfo{pages}{085116} (\bibinfo{year}{2015}).

\bibitem[{\citenamefont{Scheurer et~al.}(2018)\citenamefont{Scheurer,
  Chatterjee, Wu, Ferrero, Georges, and Sachdev}}]{scheurer2018}
\bibinfo{author}{\bibfnamefont{M.~S.} \bibnamefont{Scheurer}},
  \bibinfo{author}{\bibfnamefont{S.}~\bibnamefont{Chatterjee}},
  \bibinfo{author}{\bibfnamefont{W.}~\bibnamefont{Wu}},
  \bibinfo{author}{\bibfnamefont{M.}~\bibnamefont{Ferrero}},
  \bibinfo{author}{\bibfnamefont{A.}~\bibnamefont{Georges}}, \bibnamefont{and}
  \bibinfo{author}{\bibfnamefont{S.}~\bibnamefont{Sachdev}},
  \bibinfo{journal}{PNAS}
  \textbf{\bibinfo{volume}{115}}, \bibinfo{pages}{E3665}
  (\bibinfo{year}{2018}).

\bibitem[{\citenamefont{Huang and Devereaux}(2019)}]{huang2019}
\bibinfo{author}{\bibfnamefont{E.}~\bibnamefont{Huang}} \bibnamefont{and}
  \bibinfo{author}{\bibfnamefont{T.}~\bibnamefont{Devereaux}}
  (\bibinfo{year}{2019}), \bibinfo{note}{unpublished}.

\bibitem[{\citenamefont{Xu et~al.}(2009)\citenamefont{Xu, Gu, H{\"u}cker,
  Fauqu{\'e}, Perring, Regnault, and Tranquada}}]{Xu2009}
\bibinfo{author}{\bibfnamefont{G.}~\bibnamefont{Xu}},
  \bibinfo{author}{\bibfnamefont{G.~D.} \bibnamefont{Gu}},
  \bibinfo{author}{\bibfnamefont{M.}~\bibnamefont{H{\"u}cker}},
  \bibinfo{author}{\bibfnamefont{B.}~\bibnamefont{Fauqu{\'e}}},
  \bibinfo{author}{\bibfnamefont{T.~G.} \bibnamefont{Perring}},
  \bibinfo{author}{\bibfnamefont{L.~P.} \bibnamefont{Regnault}},
  \bibnamefont{and} \bibinfo{author}{\bibfnamefont{J.~M.}
  \bibnamefont{Tranquada}}, \bibinfo{journal}{Nat. Phys.}
  \textbf{\bibinfo{volume}{5}}, \bibinfo{pages}{642 EP }.

\bibitem[{\citenamefont{Hinkov et~al.}(2008)\citenamefont{Hinkov, Haug,
  Fauqu{\'e}, Bourges, Sidis, Ivanov, Bernhard, Lin, and Keimer}}]{hinkov2008}
\bibinfo{author}{\bibfnamefont{V.}~\bibnamefont{Hinkov}},
  \bibinfo{author}{\bibfnamefont{D.}~\bibnamefont{Haug}},
  \bibinfo{author}{\bibfnamefont{B.}~\bibnamefont{Fauqu{\'e}}},
  \bibinfo{author}{\bibfnamefont{P.}~\bibnamefont{Bourges}},
  \bibinfo{author}{\bibfnamefont{Y.}~\bibnamefont{Sidis}},
  \bibinfo{author}{\bibfnamefont{A.}~\bibnamefont{Ivanov}},
  \bibinfo{author}{\bibfnamefont{C.}~\bibnamefont{Bernhard}},
  \bibinfo{author}{\bibfnamefont{C.~T.} \bibnamefont{Lin}}, \bibnamefont{and}
  \bibinfo{author}{\bibfnamefont{B.}~\bibnamefont{Keimer}},
  \bibinfo{journal}{Science} \textbf{\bibinfo{volume}{319}},
  \bibinfo{pages}{597} (\bibinfo{year}{2008}).

\bibitem[{\citenamefont{Daou et~al.}(2010)\citenamefont{Daou, Chang, LeBoeuf,
  Cyr-Choini{\`e}re, Lalibert{\'e}, Doiron-Leyraud, Ramshaw, Liang, Bonn, Hardy
  et~al.}}]{Daou2009}
\bibinfo{author}{\bibfnamefont{R.}~\bibnamefont{Daou}},
  \bibinfo{author}{\bibfnamefont{J.}~\bibnamefont{Chang}},
  \bibinfo{author}{\bibfnamefont{D.}~\bibnamefont{LeBoeuf}},
  \bibinfo{author}{\bibfnamefont{O.}~\bibnamefont{Cyr-Choini{\`e}re}},
  \bibinfo{author}{\bibfnamefont{F.}~\bibnamefont{Lalibert{\'e}}},
  \bibinfo{author}{\bibfnamefont{N.}~\bibnamefont{Doiron-Leyraud}},
  \bibinfo{author}{\bibfnamefont{B.~J.} \bibnamefont{Ramshaw}},
  \bibinfo{author}{\bibfnamefont{R.}~\bibnamefont{Liang}},
  \bibinfo{author}{\bibfnamefont{D.~A.} \bibnamefont{Bonn}},
  \bibinfo{author}{\bibfnamefont{W.~N.} \bibnamefont{Hardy}},
  \bibnamefont{et~al.}, \bibinfo{journal}{Nature}
  \textbf{\bibinfo{volume}{463}}, \bibinfo{pages}{519} (\bibinfo{year}{2010}).

\bibitem[{\citenamefont{Wu et~al.}(2015)\citenamefont{Wu, Mayaffre, Kr{\"a}mer,
  Horvati{\'c}, Berthier, Hardy, Liang, Bonn, and Julien}}]{wu2015}
\bibinfo{author}{\bibfnamefont{T.}~\bibnamefont{Wu}},
  \bibinfo{author}{\bibfnamefont{H.}~\bibnamefont{Mayaffre}},
  \bibinfo{author}{\bibfnamefont{S.}~\bibnamefont{Kr{\"a}mer}},
  \bibinfo{author}{\bibfnamefont{M.}~\bibnamefont{Horvati{\'c}}},
  \bibinfo{author}{\bibfnamefont{C.}~\bibnamefont{Berthier}},
  \bibinfo{author}{\bibfnamefont{W.~N.} \bibnamefont{Hardy}},
  \bibinfo{author}{\bibfnamefont{R.}~\bibnamefont{Liang}},
  \bibinfo{author}{\bibfnamefont{D.~A.} \bibnamefont{Bonn}}, \bibnamefont{and}
  \bibinfo{author}{\bibfnamefont{M.-H.} \bibnamefont{Julien}},
  \bibinfo{journal}{Nat. Commun.} \textbf{\bibinfo{volume}{6}},
  \bibinfo{pages}{6438} (\bibinfo{year}{2015}).

\bibitem[{\citenamefont{Parker et~al.}(2010)\citenamefont{Parker, Aynajian,
  da~Silva~Neto, Pushp, Ono, Wen, Xu, Gu, and Yazdani}}]{parker2010}
\bibinfo{author}{\bibfnamefont{C.~V.} \bibnamefont{Parker}},
  \bibinfo{author}{\bibfnamefont{P.}~\bibnamefont{Aynajian}},
  \bibinfo{author}{\bibfnamefont{E.~H.} \bibnamefont{da~Silva~Neto}},
  \bibinfo{author}{\bibfnamefont{A.}~\bibnamefont{Pushp}},
  \bibinfo{author}{\bibfnamefont{S.}~\bibnamefont{Ono}},
  \bibinfo{author}{\bibfnamefont{J.}~\bibnamefont{Wen}},
  \bibinfo{author}{\bibfnamefont{Z.}~\bibnamefont{Xu}},
  \bibinfo{author}{\bibfnamefont{G.}~\bibnamefont{Gu}}, \bibnamefont{and}
  \bibinfo{author}{\bibfnamefont{A.}~\bibnamefont{Yazdani}},
  \bibinfo{journal}{Nature} \textbf{\bibinfo{volume}{468}}, \bibinfo{pages}{677} (\bibinfo{year}{2010}).

\bibitem[{\citenamefont{Cyr-Choini\`ere
  et~al.}(2015)\citenamefont{Cyr-Choini\`ere, Grissonnanche, Badoux, Day, Bonn,
  Hardy, Liang, Doiron-Leyraud, and Taillefer}}]{cyrchoiniere2015}
\bibinfo{author}{\bibfnamefont{O.}~\bibnamefont{Cyr-Choini\`ere}},
  \bibinfo{author}{\bibfnamefont{G.}~\bibnamefont{Grissonnanche}},
  \bibinfo{author}{\bibfnamefont{S.}~\bibnamefont{Badoux}},
  \bibinfo{author}{\bibfnamefont{J.}~\bibnamefont{Day}},
  \bibinfo{author}{\bibfnamefont{D.~A.} \bibnamefont{Bonn}},
  \bibinfo{author}{\bibfnamefont{W.~N.} \bibnamefont{Hardy}},
  \bibinfo{author}{\bibfnamefont{R.}~\bibnamefont{Liang}},
  \bibinfo{author}{\bibfnamefont{N.}~\bibnamefont{Doiron-Leyraud}},
  \bibnamefont{and}
  \bibinfo{author}{\bibfnamefont{L.}~\bibnamefont{Taillefer}},
  \bibinfo{journal}{Phys. Rev. B} \textbf{\bibinfo{volume}{92}},
  \bibinfo{pages}{224502} (\bibinfo{year}{2015}).

\bibitem[{\citenamefont{Emery et~al.}(1997)\citenamefont{Emery, Kivelson, and
  Zachar}}]{emery-1997}
\bibinfo{author}{\bibfnamefont{V.~J.} \bibnamefont{Emery}},
  \bibinfo{author}{\bibfnamefont{S.~A.} \bibnamefont{Kivelson}},
  \bibnamefont{and} \bibinfo{author}{\bibfnamefont{O.}~\bibnamefont{Zachar}},
  \bibinfo{journal}{Phys. Rev. B} \textbf{\bibinfo{volume}{56}},
  \bibinfo{pages}{6120} (\bibinfo{year}{1997}).

\bibitem[{\citenamefont{{Tranquada}}(2019)}]{tranquada-2019_2}
\bibinfo{author}{\bibfnamefont{J.~M.} \bibnamefont{{Tranquada}}},
 \bibinfo{eid}{arXiv:1904.10473}
  (\bibinfo{year}{2019}).

\bibitem[{\citenamefont{Achkar et~al.}(2013)\citenamefont{Achkar, He, Sutarto,
  Geck, Zhang, Kim, and Hawthorn}}]{achkar2013}
\bibinfo{author}{\bibfnamefont{A.~J.} \bibnamefont{Achkar}},
  \bibinfo{author}{\bibfnamefont{F.}~\bibnamefont{He}},
  \bibinfo{author}{\bibfnamefont{R.}~\bibnamefont{Sutarto}},
  \bibinfo{author}{\bibfnamefont{J.}~\bibnamefont{Geck}},
  \bibinfo{author}{\bibfnamefont{H.}~\bibnamefont{Zhang}},
  \bibinfo{author}{\bibfnamefont{Y.-J.} \bibnamefont{Kim}}, \bibnamefont{and}
  \bibinfo{author}{\bibfnamefont{D.~G.} \bibnamefont{Hawthorn}},
  \bibinfo{journal}{Phys. Rev. Lett.} \textbf{\bibinfo{volume}{110}},
  \bibinfo{pages}{017001} (\bibinfo{year}{2013}).

\bibitem[{\citenamefont{{McMahon} et~al.}(2019)\citenamefont{{McMahon},
  {Achkar}, {da Silva Neto}, {Djianto}, {Menard}, {He}, {Sutarto}, {Comin},
  {Liang}, {Bonn} et~al.}}]{mcmahon2019}
\bibinfo{author}{\bibfnamefont{C.}~\bibnamefont{{McMahon}}},
  \bibinfo{author}{\bibfnamefont{A.~J.} \bibnamefont{{Achkar}}},
  \bibinfo{author}{\bibfnamefont{E.~H.} \bibnamefont{{da Silva Neto}}},
  \bibinfo{author}{\bibfnamefont{I.}~\bibnamefont{{Djianto}}},
  \bibinfo{author}{\bibfnamefont{J.}~\bibnamefont{{Menard}}},
  \bibinfo{author}{\bibfnamefont{F.}~\bibnamefont{{He}}},
  \bibinfo{author}{\bibfnamefont{R.}~\bibnamefont{{Sutarto}}},
  \bibinfo{author}{\bibfnamefont{R.}~\bibnamefont{{Comin}}},
  \bibinfo{author}{\bibfnamefont{R.}~\bibnamefont{{Liang}}},
  \bibinfo{author}{\bibfnamefont{D.~A.} \bibnamefont{{Bonn}}},
  \bibnamefont{et~al.}, \bibinfo{journal}
  \bibinfo{eid}{arXiv:1904.12929} (\bibinfo{year}{2019}).

\bibitem[{\citenamefont{Edkins et~al.}(2019)\citenamefont{Edkins, Kostin,
  Fujita, Mackenzie, Eisaki, Uchida, Sachdev, Lawler, Kim, S{\'e}amus~Davis
  et~al.}}]{edkins2019}
\bibinfo{author}{\bibfnamefont{S.~D.} \bibnamefont{Edkins}},
  \bibinfo{author}{\bibfnamefont{A.}~\bibnamefont{Kostin}},
  \bibinfo{author}{\bibfnamefont{K.}~\bibnamefont{Fujita}},
  \bibinfo{author}{\bibfnamefont{A.~P.} \bibnamefont{Mackenzie}},
  \bibinfo{author}{\bibfnamefont{H.}~\bibnamefont{Eisaki}},
  \bibinfo{author}{\bibfnamefont{S.}~\bibnamefont{Uchida}},
  \bibinfo{author}{\bibfnamefont{S.}~\bibnamefont{Sachdev}},
  \bibinfo{author}{\bibfnamefont{M.~J.} \bibnamefont{Lawler}},
  \bibinfo{author}{\bibfnamefont{E.-A.} \bibnamefont{Kim}},
  \bibinfo{author}{\bibfnamefont{J.~C.} \bibnamefont{S{\'e}amus~Davis}},
  \bibnamefont{et~al.}, \bibinfo{journal}{Science}
  \textbf{\bibinfo{volume}{364}}, \bibinfo{pages}{976} (\bibinfo{year}{2019}),
.

\bibitem[{\citenamefont{{Agterberg} et~al.}(2019)\citenamefont{{Agterberg},
  {S{\'e}amus Davis}, {Edkins}, {Fradkin}, {Van Harlingen}, {Kivelson}, {Lee},
  {Radzihovsky}, {Tranquada}, and {Wang}}}]{agterberg2019}
\bibinfo{author}{\bibfnamefont{D.~F.} \bibnamefont{{Agterberg}}},
  \bibinfo{author}{\bibfnamefont{J.~C.} \bibnamefont{{S{\'e}amus Davis}}},
  \bibinfo{author}{\bibfnamefont{S.~D.} \bibnamefont{{Edkins}}},
  \bibinfo{author}{\bibfnamefont{E.}~\bibnamefont{{Fradkin}}},
  \bibinfo{author}{\bibfnamefont{D.~J.} \bibnamefont{{Van Harlingen}}},
  \bibinfo{author}{\bibfnamefont{S.~A.} \bibnamefont{{Kivelson}}},
  \bibinfo{author}{\bibfnamefont{P.~A.} \bibnamefont{{Lee}}},
  \bibinfo{author}{\bibfnamefont{L.}~\bibnamefont{{Radzihovsky}}},
  \bibinfo{author}{\bibfnamefont{J.~M.} \bibnamefont{{Tranquada}}},
  \bibnamefont{and} \bibinfo{author}{\bibfnamefont{Y.}~\bibnamefont{{Wang}}},
\bibinfo{eid}{arXiv:1904.09687}
  (\bibinfo{year}{2019}).
  
  
\bibitem[{\citenamefont{Zhang et~al.}(2019)\citenamefont{Zhang, Mesaros,
  Fujita, Edkins, Hamidian, Ch'ng, Eisaki, Uchida, Davis, Khatami
  et~al.}}]{Zhang2019}
\bibinfo{author}{\bibfnamefont{Y.}~\bibnamefont{Zhang}},
  \bibinfo{author}{\bibfnamefont{A.}~\bibnamefont{Mesaros}},
  \bibinfo{author}{\bibfnamefont{K.}~\bibnamefont{Fujita}},
  \bibinfo{author}{\bibfnamefont{S.~D.} \bibnamefont{Edkins}},
  \bibinfo{author}{\bibfnamefont{M.~H.} \bibnamefont{Hamidian}},
  \bibinfo{author}{\bibfnamefont{K.}~\bibnamefont{Ch'ng}},
  \bibinfo{author}{\bibfnamefont{H.}~\bibnamefont{Eisaki}},
  \bibinfo{author}{\bibfnamefont{S.}~\bibnamefont{Uchida}},
  \bibinfo{author}{\bibfnamefont{J.~C.~S.} \bibnamefont{Davis}},
  \bibinfo{author}{\bibfnamefont{E.}~\bibnamefont{Khatami}},
  \bibnamefont{et~al.}, \bibinfo{journal}{Nature}
  \textbf{\bibinfo{volume}{570}}, \bibinfo{pages}{484} (\bibinfo{year}{2019}).



\bibitem[{\citenamefont{{Zhang} et~al.}(2018)\citenamefont{{Zhang}, {Mesaros},
  {Fujita}, {Edkins}, {Hamidian}, {Ch'ng}, {Eisaki}, {Uchida}, {S{\'e}amus
  Davis}, {Khatami} et~al.}}]{Zhang2019}
\bibinfo{author}{\bibfnamefont{Y.}~\bibnamefont{{Zhang}}},
  \bibinfo{author}{\bibfnamefont{A.}~\bibnamefont{{Mesaros}}},
  \bibinfo{author}{\bibfnamefont{K.}~\bibnamefont{{Fujita}}},
  \bibinfo{author}{\bibfnamefont{S.~D.} \bibnamefont{{Edkins}}},
  \bibinfo{author}{\bibfnamefont{M.~H.} \bibnamefont{{Hamidian}}},
  \bibinfo{author}{\bibfnamefont{K.}~\bibnamefont{{Ch'ng}}},
  \bibinfo{author}{\bibfnamefont{H.}~\bibnamefont{{Eisaki}}},
  \bibinfo{author}{\bibfnamefont{S.}~\bibnamefont{{Uchida}}},
  \bibinfo{author}{\bibfnamefont{J.~C.} \bibnamefont{{S{\'e}amus Davis}}},
  \bibinfo{author}{\bibfnamefont{E.}~\bibnamefont{{Khatami}}},
  \bibnamefont{et~al.},  \bibinfo{eid}{arXiv:1808.00479} (\bibinfo{year}{2018}).

\bibitem[{\citenamefont{Zheng et~al.}(2017)\citenamefont{Zheng, Chung, Corboz,
  Ehlers, Qin, Noack, Shi, White, Zhang, and Chan}}]{zheng2017}
\bibinfo{author}{\bibfnamefont{B.-X.} \bibnamefont{Zheng}},
  \bibinfo{author}{\bibfnamefont{C.-M.} \bibnamefont{Chung}},
  \bibinfo{author}{\bibfnamefont{P.}~\bibnamefont{Corboz}},
  \bibinfo{author}{\bibfnamefont{G.}~\bibnamefont{Ehlers}},
  \bibinfo{author}{\bibfnamefont{M.-P.} \bibnamefont{Qin}},
  \bibinfo{author}{\bibfnamefont{R.~M.} \bibnamefont{Noack}},
  \bibinfo{author}{\bibfnamefont{H.}~\bibnamefont{Shi}},
  \bibinfo{author}{\bibfnamefont{S.~R.} \bibnamefont{White}},
  \bibinfo{author}{\bibfnamefont{S.}~\bibnamefont{Zhang}}, \bibnamefont{and}
  \bibinfo{author}{\bibfnamefont{G.~K.-L.} \bibnamefont{Chan}},
  \bibinfo{journal}{Science} \textbf{\bibinfo{volume}{358}},
  \bibinfo{pages}{1155} (\bibinfo{year}{2017}).

\bibitem[{\citenamefont{Corboz et~al.}(2014)\citenamefont{Corboz, Rice, and
  Troyer}}]{corboz2014}
\bibinfo{author}{\bibfnamefont{P.}~\bibnamefont{Corboz}},
  \bibinfo{author}{\bibfnamefont{T.~M.} \bibnamefont{Rice}}, \bibnamefont{and}
  \bibinfo{author}{\bibfnamefont{M.}~\bibnamefont{Troyer}},
  \bibinfo{journal}{Phys. Rev. Lett.} \textbf{\bibinfo{volume}{113}},
  \bibinfo{pages}{046402} (\bibinfo{year}{2014}).

\bibitem[{\citenamefont{Cyr-Choini\`ere
  et~al.}(2018)\citenamefont{Cyr-Choini\`ere, LeBoeuf, Badoux,
  Dufour-Beaus\'ejour, Bonn, Hardy, Liang, Graf, Doiron-Leyraud, and
  Taillefer}}]{cyrchoiniere2018}
\bibinfo{author}{\bibfnamefont{O.}~\bibnamefont{Cyr-Choini\`ere}},
  \bibinfo{author}{\bibfnamefont{D.}~\bibnamefont{LeBoeuf}},
  \bibinfo{author}{\bibfnamefont{S.}~\bibnamefont{Badoux}},
  \bibinfo{author}{\bibfnamefont{S.}~\bibnamefont{Dufour-Beaus\'ejour}},
  \bibinfo{author}{\bibfnamefont{D.~A.} \bibnamefont{Bonn}},
  \bibinfo{author}{\bibfnamefont{W.~N.} \bibnamefont{Hardy}},
  \bibinfo{author}{\bibfnamefont{R.}~\bibnamefont{Liang}},
  \bibinfo{author}{\bibfnamefont{D.}~\bibnamefont{Graf}},
  \bibinfo{author}{\bibfnamefont{N.}~\bibnamefont{Doiron-Leyraud}},
  \bibnamefont{and}
  \bibinfo{author}{\bibfnamefont{L.}~\bibnamefont{Taillefer}},
  \bibinfo{journal}{Phys. Rev. B} \textbf{\bibinfo{volume}{98}},
  \bibinfo{pages}{064513} (\bibinfo{year}{2018}).

\bibitem[{\citenamefont{Brouet et~al.}(2008)\citenamefont{Brouet, Yang, Zhou,
  Hussain, Moore, He, Lu, Shen, Laverock, Dugdale et~al.}}]{Brouet2008}
\bibinfo{author}{\bibfnamefont{V.}~\bibnamefont{Brouet}},
  \bibinfo{author}{\bibfnamefont{W.~L.} \bibnamefont{Yang}},
  \bibinfo{author}{\bibfnamefont{X.~J.} \bibnamefont{Zhou}},
  \bibinfo{author}{\bibfnamefont{Z.}~\bibnamefont{Hussain}},
  \bibinfo{author}{\bibfnamefont{R.~G.} \bibnamefont{Moore}},
  \bibinfo{author}{\bibfnamefont{R.}~\bibnamefont{He}},
  \bibinfo{author}{\bibfnamefont{D.~H.} \bibnamefont{Lu}},
  \bibinfo{author}{\bibfnamefont{Z.~X.} \bibnamefont{Shen}},
  \bibinfo{author}{\bibfnamefont{J.}~\bibnamefont{Laverock}},
  \bibinfo{author}{\bibfnamefont{S.~B.} \bibnamefont{Dugdale}},
  \bibnamefont{et~al.}, \bibinfo{journal}{Phys. Rev. B}
  \textbf{\bibinfo{volume}{77}}, \bibinfo{pages}{235104}
  (\bibinfo{year}{2008}).

\bibitem[{\citenamefont{Matt et~al.}(2015)\citenamefont{Matt, Fatuzzo, Sassa,
  Mansson, Fatale, Bitetta, Shi, Pailh\`es, Berntsen, Kurosawa
  et~al.}}]{Matt2015}
\bibinfo{author}{\bibfnamefont{C.~E.} \bibnamefont{Matt}},
  \bibinfo{author}{\bibfnamefont{C.~G.} \bibnamefont{Fatuzzo}},
  \bibinfo{author}{\bibfnamefont{Y.}~\bibnamefont{Sassa}},
  \bibinfo{author}{\bibfnamefont{M.}~\bibnamefont{Mansson}},
  \bibinfo{author}{\bibfnamefont{S.}~\bibnamefont{Fatale}},
  \bibinfo{author}{\bibfnamefont{V.}~\bibnamefont{Bitetta}},
  \bibinfo{author}{\bibfnamefont{X.}~\bibnamefont{Shi}},
  \bibinfo{author}{\bibfnamefont{S.}~\bibnamefont{Pailh\`es}},
  \bibinfo{author}{\bibfnamefont{M.~H.} \bibnamefont{Berntsen}},
  \bibinfo{author}{\bibfnamefont{T.}~\bibnamefont{Kurosawa}},
  \bibnamefont{et~al.}, \bibinfo{journal}{Phys. Rev. B}
  \textbf{\bibinfo{volume}{92}}, \bibinfo{pages}{134524}
  (\bibinfo{year}{2015}).

\bibitem[{\citenamefont{Wu et~al.}(2011)\citenamefont{Wu, Mayaffre, Kr{\"a}mer,
  Horvati{\'c}, Berthier, Hardy, Liang, Bonn, and Julien}}]{Wu2011}
\bibinfo{author}{\bibfnamefont{T.}~\bibnamefont{Wu}},
  \bibinfo{author}{\bibfnamefont{H.}~\bibnamefont{Mayaffre}},
  \bibinfo{author}{\bibfnamefont{S.}~\bibnamefont{Kr{\"a}mer}},
  \bibinfo{author}{\bibfnamefont{M.}~\bibnamefont{Horvati{\'c}}},
  \bibinfo{author}{\bibfnamefont{C.}~\bibnamefont{Berthier}},
  \bibinfo{author}{\bibfnamefont{W.~N.} \bibnamefont{Hardy}},
  \bibinfo{author}{\bibfnamefont{R.}~\bibnamefont{Liang}},
  \bibinfo{author}{\bibfnamefont{D.~A.} \bibnamefont{Bonn}}, \bibnamefont{and}
  \bibinfo{author}{\bibfnamefont{M.-H.} \bibnamefont{Julien}},
  \bibinfo{journal}{Nature} \textbf{\bibinfo{volume}{477}}, \bibinfo{pages}{} (\bibinfo{year}{2011}).

\bibitem[{\citenamefont{Kivelson and Kivelson}(2018)}]{Kivelson2018}
\bibinfo{author}{\bibfnamefont{S.}~\bibnamefont{Kivelson}} \bibnamefont{and}
  \bibinfo{author}{\bibfnamefont{S.}~\bibnamefont{Kivelson}},
  \bibinfo{journal}{Nat. Phys.} \textbf{\bibinfo{volume}{14}},
  \bibinfo{pages}{426} (\bibinfo{year}{2018}).

\end{thebibliography}
\end{document}